# A combined immersed boundary and discrete unified gas kinetic scheme for particle-fluid flows

Shi Tao[1], Haolong Zhang[1], Zhaoli Guo[1,2,*], Lian-Ping Wang[1,3]

[1] State Key Laboratory of coal Combustion, School of Energy and Power Engineering, Huazhong University of Science and Technology, Wuhan 430074, Hubei, China

[2] Beijing Computational Science Research Center, Beijing 100084, China.

[3] Department of Mechanical Engineering, University of Delaware, Newark, DE 19716-3140, USA

**Abstract**

A discrete unified gas kinetic scheme (DUGKS) coupled with the immersed boundary (IB) method is developed to perform interface-resolved simulation of particle-laden flows. The present method (IB-DUGKS) preserves the respective advantages of the IB and DUGKS, i.e., the flexibility and efficiency for treating complex flows, and the robustness and low numerical-dissipation. In IB-DUGKS, the IB method is used to treat the fluid-solid interfaces and the DUGKS is applied to simulate the fluid flow, making use of the Lagrangian and Eulerian meshes, respectively. Those two meshes are fully independent, which contributes to the avoidance of grid regeneration when a solid particle moves. Specifically, in the present implementation of IB-DUGKS, the no-slip boundary condition at the particle surface is accurately enforced by introducing an efficient iterative forcing algorithm, and the IB force induced by the particle boundary is conveniently incorporated into the DUGKS with the Strange-Splitting scheme. The accuracy of the IB-DUGKS is first verified in the flows past a stationary cylinder and an oscillating cylinder in a quiescent fluid. After that, several well-established two- and three-dimensional particulate flow problems are simulated, including the sedimentation of a particle and the DKT dynamics of two particles in a channel, and a group of particles settling in an enclosure. In all test cases, the results are in good agreement with the data available in the literature, demonstrating that the proposed IB-DUGKS is a promising tool for simulating particulate flows.

## 1. Introduction

Particulate flows are encountered in a wide range of natural and industrial processes, such as paper manufacturing, bed fluidization, oil cracking and cell sorting [1,2]. With the growing demand for better system design and control in these applications, a quite challenging but essential task is to obtain a deep understanding of the particle-fluid and particle-particle interactions in these flows. While the Eulerian-Eulerian and Eulerian-Lagragian models based on the point-particle representation have gained much success in predicting the macroscopic feature of the particulate flows [2,17], such representation is not always suitable. In general, the hydrodynamic force acting on a finite-size solid particle in a non-uniform flow remains an open fundamental question, in particular for dense and inhomogeneous particle-laden flows. It is also a non-trivial task to accurately consider the influence of finite-size particles on the dynamics of carrier fluid, their mutual interactions and interactions with a solid wall, arising in problems involving particles aggregation and deposition. On the other hand, these interactions can be well accounted in the particle-resolved direct numerical simulation (PR-DNS) [3], in which each particle is resolved by a few numerical grids, and the hydrodynamic force coming from fluid can be obtained from the simulated fluid flow observing the no-slip boundary condition on the surface of the moving solid particle.

---

*zlguo@hust.edu.cn

In PR-DNS, there are mainly two strategies available in literature to handle the moving solid-fluid interfaces. One is the arbitrary Lagrange- Euler (ALE) method [4], which uses a body-fitted grid and regenerates the computation mesh whenever a particle moves (i.e., moving mesh). Although the accuracy of the no-slip boundary treatment can be guaranteed in the ALE method, the computational cost is usually quite expensive for managing an evolving adaptive mesh at each time step. This problem becomes more serious for three-dimensional problems with multiple particles. The second type of PR-DNS approach uses a fixed Cartesian grid over which the particle-fluid interfaces move, for example, the fictitious domain method (FDM) [5], lattice Boltzmann method (LBM) [6] and immersed boundary method (IBM) [7]. Since the time-consuming re-meshing is avoided, such methods are computationally more efficient, making the investigation with a large number of particles possible, and thus have received more and more attentions in the field of particulate flows [8-10,39].

The present work makes use of the IBM [7,11] for its simplicity in programming and flexibility in applications. In IBM, an Eulerian mesh is used for simulating the fluid flow which fills the entire domain including the interior of particles. For each solid particle, its surface is discretized into a set of Lagrngian points, which serves as a source of external force that will be distributed to the ambient Eulerian grids. Hence, a force term is to be added to the momentum equation of fluid. In this way, the fluid feels the existence of the immersed boundary representing the effect of the solid particle, to properly realize the kinematic velocity no-slip boundary condition. The key point in IBM is the determination of the IB force. Several such force-coupling schemes have been proposed in the literature, such as the spring force [12], feedback force [13] and the direct force [14] models. Compared to the former two, the direct force (DF) method is widely used, due to the advantages that no empirical parameters are introduced and no additional constraint is imposed for the time step size. In DF scheme, the IB force is calculated according to the velocity difference between the two types of meshes at the Lagrangian points. Therefore, the information of the Eulerian grids is needed in this procedure, and it is usually obtained using a smoothed $\delta$-function [15]. Note that such same function is used to transfer the coupling force back to the Eulerian grids.

The IBM is generally used for the treatment of boundary of a body. Hence, an efficient flow solver is still needed to simulate the flow of the carrier fluid. In most cases in the past, IBM was combined with the Navier-Stokes (N-S) solvers. Recently, some efforts were also made to combine IBM with the lattice Boltzmann method [12,18,55]. Compared to the traditional N-S solvers, LBM is a kinetic method. Hence, there is no need for LBM to solve the time-consuming Poisson equation, since the pressure in LBM is determined directly by the equation of state. In addition, LBM has some other distinctive merits such as easy implementation and natural parallelism. For those reasons, the application of LBM as an alternative flow solver has been advanced dramatically in recent years [19,34]. However, several weaknesses of LBM have also been unveiled, for instant the time step-space lattice coupling severely restricts the form of computational grid [21,49]. Meanwhile, it is noted that some other kinetic methods, such as the gas kinetic scheme (GKS) [22], lattice Boltzmann flux solver (LBFS) [16], and gas kinetic flux solver (GKFS) [23] have also been combined with the IBM.

Most recently, a new kinetic method called DUGKS (discrete unified gas kinetic scheme) [20,24] was proposed that combines the advantages of LBM and GKS. This scheme is formulated based on the finite volume method applied directly to the Boltzmann equation. Hence, the limitation in LBM to simple rectangular grid no longer exists in DUGKS [26]. Furthermore, the temporal and spatial steps are fully decoupled, and thus the temporal step is determined by the Courant-Friedrichs-Lewy (CFL) number. As an intrinsic multi-scale approach, successful applications for flows from continuum to free molecular regimes have been achieved [25,26]. As a relatively new kinetic method, the application of DUGKS for complex hydrodynamic flows such as particle-laden flow has not yet been explored.

The present work is motivated by the desire to develop a new PR-DNS approach by combining the IBM and DUGKS. The remaining part of this paper is organized as follows. A mathematical formulation of the IB-DUGKS is first presented in Sec. II, with emphases on the Strang-Splitting algorithm for including the IB force and the iterative forcing strategy for enforcing the no-slip boundary condition on the solid-fluid interfaces. Section III is devoted to validating the present method, followed with simulations of several well-established two- and three-dimensional particulate flows, including the sedimentation of particle in a vertical channel, the drafting-kissing-tumbling (DKT) dynamics of two particles and a group of particles settling in an enclosure. Finally, conclusions are drawn in Sec. IV.

## 2. Numerical methodology

### 2.1. Discrete unified gas kinetic method

The isothermal viscous flow are governed by the following N-S equations,

$$\frac{\partial \rho}{\partial t} + \nabla \cdot (\rho \boldsymbol{u}) = 0, \tag{1}$$

$$\frac{\partial}{\partial t}(\rho \boldsymbol{u}) + \nabla \cdot (\rho \boldsymbol{u}\boldsymbol{u}) = -\nabla p + \nabla \cdot \left[ \mu \left( \nabla \boldsymbol{u} + \nabla \boldsymbol{u}^T \right) \right], \tag{2}$$

where $\rho$, $\boldsymbol{u}$, $p$ and $\mu$ are the fluid density, velocity, pressure and dynamic viscosity, respectively. Many numerical approaches have been developed to solve the N-S equations, such as the classical CFD methods [27], LBM [19] and DUGKS [20]. The recently proposed DUGKS is based on the Boltzmann equation and designed virtually for multi-scale flow problems. The application of DUGKS for the incompressible viscous flows has been extended and well demonstrated subsequently [25]. A detailed description of the method can be found in [20,24], and a brief introduction is given below.

The starting point of the DUGKS is the Boltzmann equation with the Bhatnagar -Gross-Krook (BGK) collision model,

$$\frac{\partial f}{\partial t} + \boldsymbol{\xi} \cdot \nabla f = \Omega = \frac{f^{eq} - f}{\tau}, \tag{3}$$

which describes the evolution of particle distribution function $f = f(\boldsymbol{x}, \boldsymbol{\xi}, t)$ with velocity $\boldsymbol{\xi}$ at position $\boldsymbol{x}$ and time $t$. The collision term $\Omega$ is approximated by the BGK model, with $\tau$ being the relaxation time and $f^{eq}$ the Maxwellian equilibrium distribution function

$$f^{eq} = \frac{\rho}{(2\pi RT)^{D/2}} \exp\left(-\frac{(\boldsymbol{\xi}-\boldsymbol{u})^2}{2RT}\right), \tag{4}$$

where $R$, $T$, and $D$ are the gas constant, temperature, and spatial dimension, respectively. The conservative flow variables ($\rho$ and $\boldsymbol{u}$) are defined as the moments of the distribution function

$$\rho = \int f d\boldsymbol{\xi}, \qquad \rho \boldsymbol{u} = \int \boldsymbol{\xi} f d\boldsymbol{\xi}. \tag{5}$$

With the midpoint rule for the convection term and the trapezoidal scheme for the collision term, the DUGKS formulates a discrete form of the Boltzmann equation (3), in both time and space as [20]

$$f_j^{n+1} - f_j^n + \frac{\Delta t}{|V_j|} F^{n+1/2} = \frac{\Delta t}{2}\left(\Omega_j^{n+1} + \Omega_j^n\right). \tag{6}$$

where $V_j$ is the $j$-th control volume (cell), $\Delta t$ is the time step, and

$$F^{n+1/2} = \int_{\partial V_j} (\boldsymbol{\xi}\cdot\boldsymbol{n}) f(\boldsymbol{x}, t_{n+1/2}) dS. \tag{7}$$

is the microflux across the cell interface with the outward unit vector normal $\boldsymbol{n}$. Here $f_j^n$ and $\Omega_j^n$ are the volume-averaged values in cell $V_j$ which has a volume of $|V_j|$ and surface $\partial V_j$. The update of $f_j$ in Eq. (5) is implicit due to the existence of term $\Omega_j^{n+1}$. To convert implicit form to an explicit version of DUGKS, two new distribution functions are introduced,

$$\tilde{f} = f - \frac{\Delta t}{2}\Omega, \quad \tilde{f}^+ = f + \frac{\Delta t}{2}\Omega. \tag{8}$$

Note that the following relationship holds

$$\tilde{f}^+ = \frac{2\tau - \Delta t}{2\tau + \Delta t}\tilde{f} + \frac{2\Delta t}{2\tau + \Delta t} f^{eq}. \tag{9}$$

The conserved variables can be directly computed from $\tilde{f}$ as that of $f$ given in Eq. (5) since the collision operator conserves mass and momentum. With these facts, evolution equation of DUGKS (6) can be rewritten as

$$\tilde{f}_j^{n+1} = \tilde{f}_j^{+,n} - \frac{\Delta t}{|V_j|} F^{n+1/2}. \tag{10}$$

We can then explicitly update $\tilde{f}$ instead of the original one in actual computation.

Now that the scheme is made explicit by transformation, the next task is to evaluate the flux $F^{n+1/2}$, namely, obtain the distribution function at the cell interface at time ($t_n + \Delta t/2$). To this end, we integrate the Boltzmann equation (3) along the characteristic line with a time step $h = \Delta t/2$,

$$f(\boldsymbol{x}_b, \boldsymbol{\xi}, t_n + h) - f(\boldsymbol{x}_b - \boldsymbol{\xi}h, \boldsymbol{\xi}, t_n) = \frac{h}{2}[\Omega(\boldsymbol{x}_b, \boldsymbol{\xi}, t_n + h) + \Omega(\boldsymbol{x}_b - \boldsymbol{\xi}h, \boldsymbol{\xi}, t_n)], \tag{11}$$

where $\boldsymbol{x}_b$ is a point located at the cell interface, and the trapezoidal rule is again used for the collision term. Analogous to the treatment of Eq. (6), we introduce another two auxiliary distribution functions to make Eq. (11) explicit,

$$\bar{f} = f - \frac{h}{2}\Omega, \quad \bar{f}^+ = f + \frac{h}{2}\Omega, \tag{12}$$

with

$$\bar{f}^+ = \frac{2\tau - h}{2\tau + h}\bar{f} + \frac{2h}{2\tau + h} f^{eq}. \tag{13}$$

Then Eq. (11) can be simplified to

$$\bar{f}(\boldsymbol{x}_b, \boldsymbol{\xi}, t_n + h) = \bar{f}^+(\boldsymbol{x}_b - \boldsymbol{\xi}h, \boldsymbol{\xi}, t_n)., \tag{14}$$

where the distribution function on the right hand side can be reconstructed as

$$\bar{f}(\boldsymbol{x}_b, \boldsymbol{\xi}, t_n + h) = \bar{f}^+(\boldsymbol{x}_b, \boldsymbol{\xi}, t_n) - \boldsymbol{\xi}h\cdot\nabla\bar{f}^+(\boldsymbol{x}_b, \boldsymbol{\xi}, t_n). \tag{15}$$

Similarly, the conserved variables can be computed from $\bar{f}(\boldsymbol{x}_b, \boldsymbol{\xi}, t_n + h)$ as that of $f$ given in Eq. (5), and then the equilibrium function at the cell interface can be determined. As such, the distribution function at the cell interface at time $t_n + h$ can be obtained,

$$f(\pmb{x}_b,\pmb{\xi},t_n+h) = \frac{2\tau}{2\tau+h}\overline{f}(\pmb{x}_b,\pmb{\xi},t_n+h) + \frac{h}{2\tau+h} f^{eq}(\pmb{x}_b,\pmb{\xi},t_n+h). \quad (16)$$

The microflux $F^{n+1/2}$ is then computed according to Eq. (7), and the evolution of $\tilde{f}$ can be processed according to Eq. (10). It is noted that the auxiliary functions, $\tilde{f}, \tilde{f}^+, \bar{f}$ and $\bar{f}^+$ are all related to $f$ and $f^{eq}$. In particular, the following two relations will be used in computation [20],

$$\overline{f}^+ = \frac{2\tau-h}{2\tau+\Delta t} f + \frac{3h}{2\tau+t} f^{eq}, \quad (17)$$

$$f^+ = \frac{4}{3}\overline{f}^+ - \frac{1}{3}f \quad (18)$$

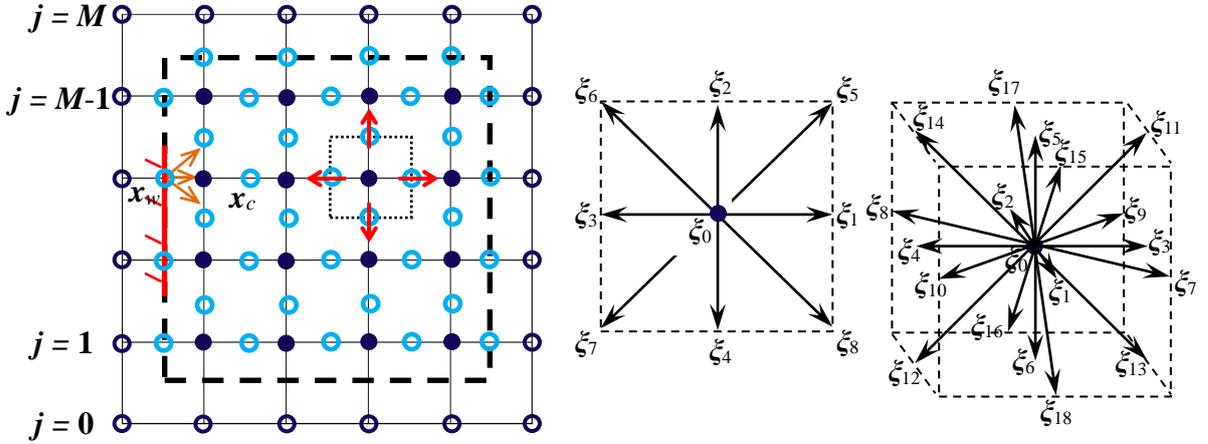

**Fig. 1.** The layout of the grids, the discrete velocity models (2D and 3D) and the straight boundary treatment in the IB-DUGKS.

By now, the discretization of the Boltzmann equation (3) with the BGK model in time and space domains is completed. As a discrete ordinate method, the remaining task in DUGKS is to discretize the velocity space. The three-point Gauss-Hermite quadrature and the tensor product method [20,28] are use in the present work to specify a minimum set of the discrete particle velocities and corresponding weights for one and higher dimension problems. For 2D and 3D flows considered in this study, the following nine- and nineteen-velocity models (D2Q9 and D3Q19), shown in Fig. 1, will be employed respectively,

$$\pmb{\xi}_i = \begin{cases} (0,0), \ i=0, \\ (\pm 1,0)c, (0,\pm 1)c, \ i=1-4, \\ (\pm 1,\pm 1)c, \ i=5-8, \end{cases} \quad (19a)$$

$$\pmb{\xi}_i = \begin{cases} (0,0), \ i=0, \\ (\pm 1,0,0)c, (0,\pm 1,0)c, (0,0,\pm 1)c, \ i=1-6, \\ (\pm 1,\pm 1,0)c, (\pm 1,0,\pm 1)c, (0,\pm 1,\pm 1)c, \ i=7-18, \end{cases} \quad (19b)$$

where $c = \sqrt{3RT}$ is the model speed of sound. The associated weights are $W_0 = 4/9$, $W_{1,2,3,4} = 1/9$, $W_{5,6,7,8} = 1/36$ for D2Q9, and $W_0 = 1/3$, $W_{1,\ldots,6} = 1/18$, $W_{7,\ldots,18} = 1/36$ for D3Q19. Under the assumption of low Mach number $\text{Ma} \approx |\pmb{u}|/c_s = |\pmb{u}|/\sqrt{RT} \ll 1.0$, the discrete equilibrium distribution function (4) can be approximated up to the second order in Ma as

$$f_i^{eq} = W_i \rho \left[ 1 + \frac{\boldsymbol{\xi}_i.\boldsymbol{u}}{RT} + \frac{(\boldsymbol{\xi}_i.\boldsymbol{u})^2}{2(RT)^2} - \frac{|\boldsymbol{u}|^2}{2RT} \right]. \tag{20}$$

The macroscopic quantities, $\rho$, $\boldsymbol{u}$, $p$ and $\mu$ are obtained as

$$\rho = \sum_i f_i, \quad \rho\boldsymbol{u} = \sum_i \boldsymbol{\xi}_i f_i, \quad p = \rho RT, \quad \mu = \rho\tau RT \tag{21}$$

Now we discuss the boundary condition for DUGKS. As shown in Fig. 1, in DUGKS a straight boundary will be located at the cell interface. For the treatment of boundary condition (BC) of a solid wall, the bounce-back (BB) rule can be used to give the distribution function pointing towards the flow field as

$$f(\boldsymbol{x}_w, \boldsymbol{\xi}_i, t+h) = f(\boldsymbol{x}_w, -\boldsymbol{\xi}_i, t+h) + 2\rho_w W_i \frac{\boldsymbol{\xi}_i.\boldsymbol{u}_w}{RT}, \tag{22}$$

where $\boldsymbol{u}_w$ is the velocity of wall, and $\rho_w$ is the fluid density near the wall which can be approximated using the average density of fluid. For open BCs, such as the channel inlet or outlet the non-equilibrium extrapolation method [50] is applied,

$$\begin{aligned} f(\boldsymbol{x}_w, \boldsymbol{\xi}_i, t+h) &= f^{eq}(\boldsymbol{\xi}_i, \rho_w, \boldsymbol{u}_w) + f^{neq}(\boldsymbol{x}_w, \boldsymbol{\xi}_i, t+h) \\ &= f^{eq}(\boldsymbol{\xi}_i, \rho_w, \boldsymbol{u}_w) + f(\boldsymbol{x}_c, \boldsymbol{\xi}_i, t+h) - f^{eq}(\boldsymbol{\xi}_i, \rho_c, \boldsymbol{u}_c). \end{aligned} \tag{23}$$

If the flow is periodic in one direction (vertical shown in Fig. 1), the distribution function at the center of a ghost cell ($j = 0$ and $M$) is determined as

$$f(\boldsymbol{x}_{j=0}, \boldsymbol{\xi}_i, t+h) = f(\boldsymbol{x}_{j=M-1}, \boldsymbol{\xi}_i, t+h), \quad f(\boldsymbol{x}_{j=M}, \boldsymbol{\xi}_i, t+h) = f(\boldsymbol{x}_{j=1}, \boldsymbol{\xi}_i, t+h). \tag{24}$$

*2.1.1. Strang-Splitting method for body force*

The framework of DUGKS presented above is for macroscopic viscous flows where no external force is involved. In some situations, a body force will be inevitably encountered, for instant the channel flow driven by an external force, or the boundary force using the IB method. Therefore, a forcing term should be added into the governing equations as

$$\frac{\partial f}{\partial t} + \boldsymbol{\xi}.\nabla f = \frac{f^{eq} - f}{\tau} + \boldsymbol{F}, \tag{25}$$

where

$$\boldsymbol{F} = -\boldsymbol{a}.\nabla_{\boldsymbol{\xi}} f \approx -\boldsymbol{a}.\nabla_{\boldsymbol{\xi}} f^{eq} = \boldsymbol{a}.\frac{(\boldsymbol{\xi}-\boldsymbol{u})}{RT} f^{eq}, \tag{26}$$

with $\boldsymbol{a}$ the acceleration due to the body force. Wu et al. [29] considered the force by modifying the equilibrium distribution function, while Yuan et al. [22] put forward a special iterative procedure that the force had an effect on the interface flux. Here, a simpler way using the Strang-Splitting algorithm [30,31] is introduced to include the force term into the Boltzmann equation (3). In the Strang-Splitting scheme, the force is to split in half, and is added before and after the DUGKS procedure as

$$\frac{\partial f}{\partial t} = \frac{1}{2}\boldsymbol{F}, \tag{27}$$

$$\frac{\partial f}{\partial t} + \boldsymbol{\xi}.\nabla f = \frac{f^{eq} - f}{\tau}, \tag{28}$$

$$\frac{\partial f}{\partial t} = \frac{1}{2}\boldsymbol{F}, \tag{29}$$

which are called as pre-forcing, DUGKS with no-forcing, and post-forcing steps, respectively. Within each forcing step (taking the pre-forcing step for example), the original distribution function as well as the macroscopic quantities are calculated as

$$f^* = f^n + 0.5\Delta t \mathbf{F}(\rho^n, \mathbf{u}^n), \tag{30}$$

$$\rho^* = \rho^n, \quad \mathbf{u}^* = \mathbf{u}^n + 0.5\mathbf{a}\Delta t \tag{31}$$

In terms of $\tilde{f}$ actually tracked in the DUGKS, it can be shown that this is equivalent to the following updating procedure

$$\overline{f}^* = \overline{f}^n + 0.5\Delta t \mathbf{F}(\rho^n, \mathbf{u}^n). \tag{32}$$

The details of the derivation of Eq. (32) can be found in the Appendix.

## 2.2. Immersed boundary method

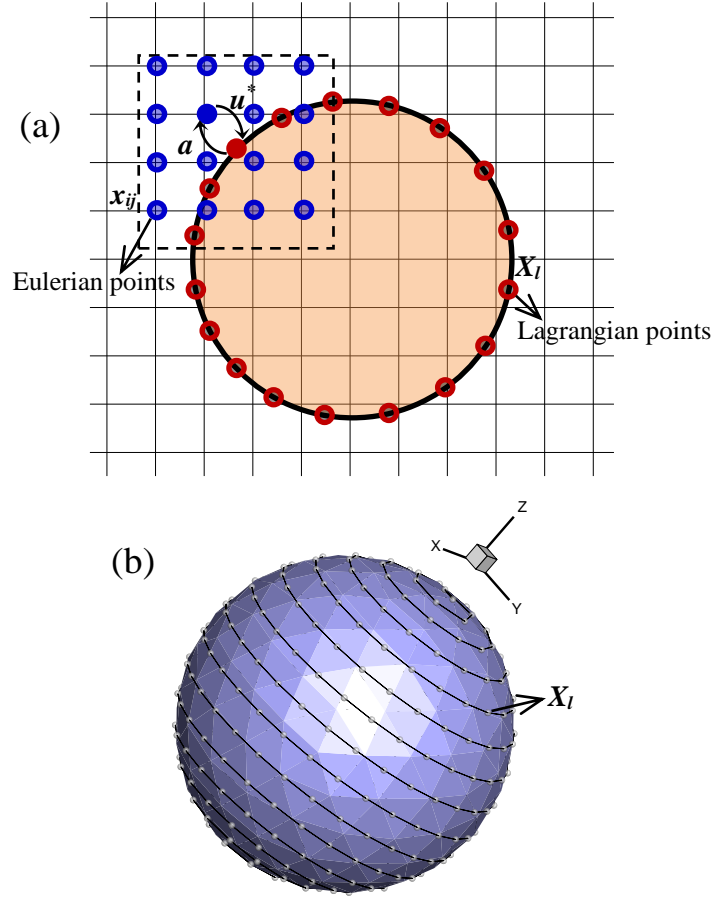

**Fig. 2.** Immersed boundary scheme for the treatment of particle boundary (a). The distribution of the Lagrangian points on the sphere (b).

The immersed boundary method is a flexible approach for the treatment of moving boundaries with complex geometry [7,15]. In IBM, the boundary of a body is discretized into a set of Lagrangian points $X_l$, as shown in Fig. 2. In general, those points should be distributed uniformly at the body boundary [7]. It is easy to achieve this in the case of a two-dimensional circular cylinder. As for a sphere, Saff et al.

[52] proposed the following method to determine the positions of the Lagrangian points as

$$\theta_k = \arccos(a_k), \quad a_k = -1 + \frac{2(k-1)}{N-1}, \quad 1 \le k \le N,$$
$$\varphi_1 = \varphi_N = 0, \quad \varphi_k = \varphi_{k-1} + \frac{3.6}{\sqrt{N(1-a_k^2)}}, \quad 1 < k < N, \tag{33}$$

where $\theta$ and $\varphi$ are the spherical coordinates. $N$ is the total number of points, which value is assumed to be no less than $\pi(D^2/\Delta x^2 + 1/3)$ [7]. Each such point serves as a source of body force acting on the fluid, which will be distributed to the ambient Eulerian points $x_{ij}$. In this way, the fluid senses the existence of the immersed boundary and the kinetic velocity boundary condition is transformed into a force field. It should be noted that the Eulerian mesh is fixed in the entire computation process.

The value of force must be determined first in the application of IBM. Several types of forcing method have been proposed in literature [12-14]. The widely-used direct forcing model will be adopted in this study, in which the IB force is given by

$$a_l = 2\frac{U - u^*}{\Delta t}, \tag{34}$$

where the $U$ and $u^*$ are the desired velocity and the fluid velocity immediately before the forcing implementation at the position of Lagrangian point $X_l$. The $u^*$ can be interpolated by the smoothed Dirac delta function as

$$D_2(\mathbf{x}_{ij} - \mathbf{X}_l) = \delta\left(\frac{x_i - x_l}{dx}\right)\delta\left(\frac{y_j - y_l}{dx}\right), \quad D_3(\mathbf{x}_{ijk} - \mathbf{X}_l) = \delta\left(\frac{x_i - x_l}{dx}\right)\delta\left(\frac{y_j - y_l}{dx}\right)\delta\left(\frac{z_k - z_l}{dx}\right),$$
$$\delta(r) = \begin{cases} \frac{1}{4}\left(1 + \cos\left(\frac{\pi|r|}{2}\right)\right), & |r| \le 2 \\ 0, & |r| > 2 \end{cases} \tag{35}$$

$$u^* = \sum_{i,j} u_{ij} D_2(\mathbf{x}_{ij} - \mathbf{X}_l) \text{ or } \sum_{i,j,k} u_{ijk} D_3(\mathbf{x}_{ijk} - \mathbf{X}_l). \tag{36}$$

The calculated force $a_l$ is then distributed back to the Eulerian points using the same delta function,

$$a = \sum_l a_l D_2(\mathbf{x}_{ij} - \mathbf{X}_l) \text{ or } \sum_l a_l D_3(\mathbf{x}_{ijk} - \mathbf{X}_l). \tag{37}$$

Using this distributed force, the velocity at the Eulerian point is updated as

$$u = u^* + 0.5\Delta t a. \tag{38}$$

*2.2.1. Iterative forcing algorithm for no-slip condition*

In some recent studies [16,23], it is found that the no-slip boundary condition is only approximately satisfied using the above-mentioned standard IBM, because the interpolated velocity $u$ after forcing still deviates from the expected velocity $U$. This can result in the phenomenon of unphysical streamline penetration, which may cause the collapse of mass conservation [32]. To avoid this defect, an iterative forcing scheme [33] will be introduced in the present study. The concept of this algorithm is simple: since the unsatisfactory velocity difference exists if the force is implemented only once, we instead apply the forcing procedure multiple times, and then each time the updated velocity $u$ at the Lagrangian point approaches gradually to the expected velocity $U$. Within a finite number of forcing ($N_F$), the error in velocity can reduce to an acceptable level which will be discussed in Sec. III. A.

## 2.3. Particle dynamics

To update the position of a freely moving particle, the total force and torque exerted on it have to be calculated. In IBM, the hydrodynamic force coming from the fluid can be simply obtained as a counter-acting force. However, it is found in the previous studies [35,51] that such method only holds for the inertial motion of a body. For the case where the object is accelerated, terms that account for the effect of inertial mass should be added into the force and torque calculation as

$$\bm{F}_s = -\sum_1^{NF}\sum_l 2\rho_f \frac{\bm{U}-\bm{u}^*}{\Delta t} + \rho_f V_s \frac{d\bm{U}_s}{dt}, \quad (39)$$

$$\bm{T}_s = -\sum_1^{NF}\sum_l \left(2\rho_f \frac{\bm{U}-\bm{u}^*}{\Delta t} \times (\bm{X}_l - \bm{X}_c)\right) + \frac{\rho_f}{\rho_s} I_s \frac{d\bm{\phi}_s}{dt}, \quad (40)$$

where $V_s$, $\bm{U}_s$, $\bm{X}_c$, $\rho_s$, $I_s$ and $\bm{\phi}_s$ are the volume, velocity, mass center, density, and momentum of inertia and angular velocity of the particle, respectively. Other force that will need to be considered could include the force exerted on the particle due to the particle-particle or particle-wall collisions, which is usually given by the repulsive force scheme [12] as

$$\bm{F}_c = \begin{cases} 0, & \|\bm{x}_i - \bm{x}_j\| > R_i + R_j + \zeta, \\ \dfrac{c_{ij}}{\varepsilon_c}\left(\dfrac{\|\bm{x}_i - \bm{x}_j\| - R_i - R_j - \zeta}{\zeta}\right)^2 \left(\dfrac{\bm{x}_i - \bm{x}_j}{\|\bm{x}_i - \bm{x}_j\|}\right), & \|\bm{x}_i - \bm{x}_j\| \le R_i + R_j + \zeta. \end{cases} \quad (41)$$

Here, $c_{ij}$ is the force scale defined as the buoyancy force in suspension flows; $\varepsilon_c$ is the stiffness parameter for collisions and is set to be 0.01 in the present study; $R_i$ and $R_j$ are the radii of the two particles centered at $\bm{x}_i$ and $\bm{x}_j$, respectively; $\zeta$ is the threshold gap and takes a value of $0.05D$ ($D$ is the particle diameter). As for particle-wall collision, $\bm{x}_j$ is the position of a fictitious particle which is located symmetrically on the other side of the wall with $R_j = R_i$. It is worth mentioning that the collision force always points to the center of particle and hence it does not contribute to the torque. After the force and torque exerted on the particle are obtained, the trajectory can then be tracked by the Newton's Second Law,

$$M_s \frac{d\bm{u}_s}{dt} = \bm{F}_s + \bm{F}_c - M_s\left(1 - \frac{\rho_f}{\rho_s}\right)\bm{g}, \quad I_s \frac{d\bm{\phi}_s}{dt} = \bm{T}_s, \quad (42)$$

where $M_s$ is the mass of particle. The translation and rotation velocities of the particle are updated by solving Eq. (42) with the first-order Euler method,

$$\bm{u}_s^{n+1} = \bm{u}_s^n + \delta_t(\bm{F}_s + \bm{F}_e)/M_s, \quad \bm{\phi}_s^{n+1} = \bm{\phi}_s^n + \delta_t \bm{T}_s / I_s, \quad (43)$$

The particle position $\bm{x}_s$ and rotation angle $\theta$ can then be obtained as

$$\bm{x}_s^{n+1} = \bm{x}_s^n + \bm{u}_s^n \delta_t + \frac{1}{2}\delta_t^2 (\bm{F}_s + \bm{F}_e)/M_s, \quad \theta^{n+1} = \theta^n + \bm{\phi}_s^n \delta_t + \frac{1}{2}\delta_t^2 \bm{T}_s / I_s, \quad (44)$$

## 2.4. Accuracy verification of IB-DUGKS

In this subsection, we test the accuracy of the present IB-DUGKS. The DUGKS has already been numerically proved to be a fully second-order scheme in some recent studies [20,54]. Hence, it is important to verify its accuracy after combined with the IB method. Note that here we choose the cylindrical Couette flow, other than the decaying Taylor-Green vortex for conducting the numerical experiment, as the wall boundary existing physically in the former but added artificially to the latter case. In the cylindrical Couette flow problem, the fluid is confined by the inner and outer rotating cylinders, with speeds $\omega_1$, $\omega_2$, and radii $R_1$, $R_2$, respectively. This flow is subjected to the N-S equation using the cylindrical polar coordinate as

$$\frac{d^2 u_\theta}{dr^2} + \frac{d}{dr}\left(\frac{u_\theta}{r}\right) = 0, \tag{45}$$

Coupled with the following boundary conditions ($\omega_2 = 0$),

$$u_\theta\big|_{r=R_1} = \omega_1 R_1, \quad u_\theta\big|_{r=R_2} = 0, \tag{46}$$

the steady solution of the flow can be obtained as

$$u_\theta = \frac{\left(R_1^2 - \beta^2 r^2\right)\omega_1}{\left(1 - \beta^2\right)r}, \quad u_r = 0, \tag{47}$$

where $u_\theta$, $u_r$ are the velocity components, $r$ is the radial distance and $\beta = R_1/R_2$ is the radius ratio.

In the simulations, it is set that Ma = $U_1/c_s$ = 0.1, Reynolds number Re = $U_1(R_2 - R_1)/\nu$ = 10.0 ($U_1 = \omega_1 R_1$ and $\nu$ is the kinematic viscosity of fluid), CFL = 0.5, and $\beta$ = 0.3, 0.8. We calculate the $L_2$-norm to evaluate the numerical error as

$$L_2 - \text{norm} = \sqrt{\sum_{i=1}^{N}\left[(u_i - u_\theta)^2 + (v_i - u_r)^2\right]\bigg/N} \tag{48}$$

where $u_i$ and $v_i$ are the velocity components obtained by the IB-DUGKS in the cylindrical polar coordinate. The $L_2$-norm error under different grid resolutions is presented in Fig. 3. A generally first-order convergence rate can be observed for both the overall error ($N$ is the total number of fluid nodes in Eq. (48)) and the error along the velocity profile ($\theta = \pi/2$). Hence, it is then concluded that the IB-DUGKS is first-order accuracy in space, somewhat reducing the accuracy of the original DUGKS.

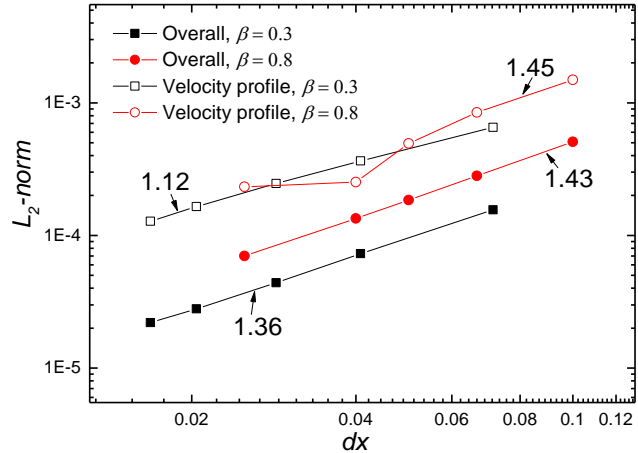

**Fig. 3.** Spatial convergence of the present IB-DUGKS in the cylindrical Couette flow.

## 3. Results and discussions

The present IB-DUGKS is validated in this section. First, the flow past a circular cylinder is simulated. As a verification of the grid-flexibility in the present method, non-uniform mesh is used in this case. Then simulations of several well-established particulate flows are investigated by the present method, including the flow by an oscillating cylinder in a quiescent fluid, the sedimentation of circular and elliptical particles in a vertical channel, the DKT dynamics of two particles and a group of particles settling in an enclosure. In the simulations, the fluid density is $\rho_f = 1.0$ g/cm$^3$, the gravity acceleration | $g$ | = 980 cm/s$^2$. For all cases in this study, the local Ma is always less than 0.3 to approximate the incompressible flows with $RT = 1/3$. The number of forcing $N_F$, unless otherwise stated is set to be 10. The number of Lagrangian points is determined by guaranteeing per arc length between two such points being half of the grid resolution. For straight wall boundary, it is handled by the bounce back rule. The non-equilibrium extrapolation is applied to the open boundary such as the channel inlet and outlet, if it is not periodic.

### 3.1. Flow past a fixed circular cylinder

The flow past a fixed circular cylinder is a classic fluid dynamic problem for which abundant numerical and experimental results are available in the literature [36-38]. The flow is controlled by Re $= U_0 D/\nu$, where $U_0$ and $D$ are the free stream velocity and cylinder diameter, respectively. For Re larger than 1 but lower than about 49, the long-time flow field is a steady state flow, and a recirculation region appears in the rear of the cylinder. Beyond the threshold, the flow could become unstable and this eventually develops the Karmen vortex shedding. The quantitative results of this problem are usually the length of recirculation zone $L_w$, separation angle $\theta$, drag force $F_d$, lift force $F_l$, and frequency of vortex shedding $f$. The latter three are respectively defined in non-dimensional form as

$$C_d = \frac{F_d}{0.5\rho_f U_0^2 D}, \quad C_l = \frac{F_l}{0.5\rho_f U_0^2 D}, \quad St = \frac{fD}{U_0}. \tag{49}$$

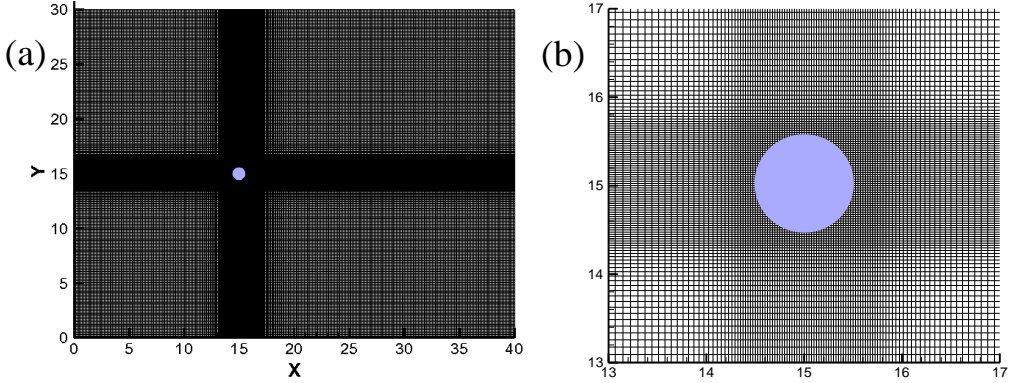

**Fig. 4.** Grid configuration used in the flow past a cylinder. (b) is the zoom-in view around the cylinder in (a).

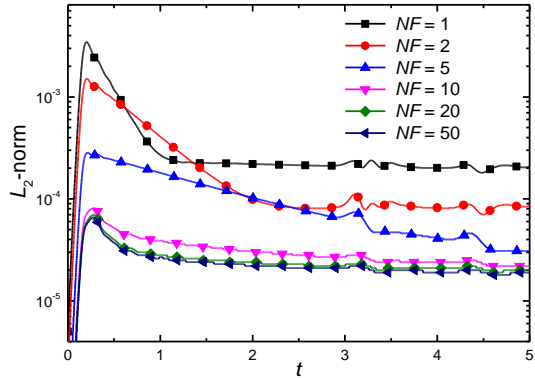

**Fig. 5.** Time history of the $L_2$-norm error in velocity at different forcing numbers.

In the simulation, size of the computational domain is (40$D$, 30$D$), and the circular cylinder is placed at the coordinate (15$D$, 15$D$). A non-uniform mesh system 399 × 308 is adopted, as shown in Fig. 4. The coarse grid resolution in the far field is $D/8$, which is refined gradually near the cylinder to be $D/40$. The computational burden is much reduced with this non-uniform mesh, in comparison with a 1600 × 1200 uniform one with resolution of $D/40$. In the simulations, the flow Ma based on $U_0$ is set as: Ma = 0.1 at Re = 20 and 40, and Ma = 0.2 at Re = 100 and 200. Constant velocity is specified at the inlet, and a free outflow is developed at the outlet. The flow is periodic in the $y$-direction.

The forcing number $N_F$ should be optimized first to best realize the no-slip boundary condition at a reasonable cost. Hence, the $L_2$-norm error in velocity at the boundary is calculated at Re = 20 with different $N_F$. The results are presented in Fig. 5. As can be seen, the errors oscillate and attenuate as time increases in all cases, implying that the no-slip boundary is accomplished gradually in IBM. Furthermore, the error decreases with increasing $N_F$. Particularly, as $N_F$ reaches 10, the error can be suppressed effectively and shows no significant differences from the results with larger values of $N_F$. Hence, we will take $N_F = 10$ in the following simulations for the sake of accuracy and efficiency.

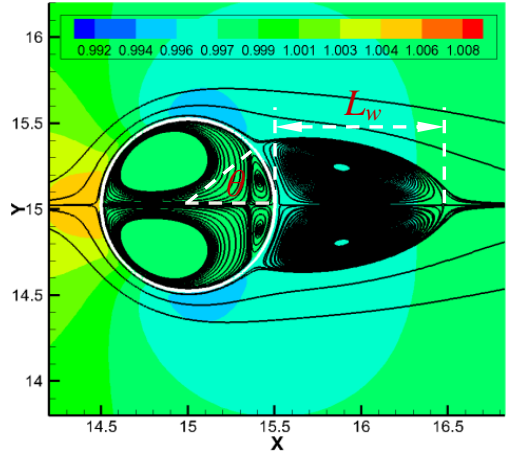

**Fig. 6.** The streamlines around and inside the cylinder at Re = 20.

TABLE 1. Comparison of the results of flow over a circular cylinder for Re = 20, 40, 100 and 200.

| Re | 20 | | | 40 | | | 100 | | | 200 | | |
|---|---|---|---|---|---|---|---|---|---|---|---|---|
| | $C_d$ | $L_w/D$ | $\theta(°)$ | $C_d$ | $L_w/D$ | $\theta(°)$ | $C_d$ | $C_l$ | $St$ | $C_d$ | $C_l$ | $St$ |
| Russell[36] | 2.17 | 0.93 | 43.9 | 1.60 | 2.29 | 53.1 | 1.43 | 0.322 | 0.172 | 1.45 | 0.63 | 0.201 |
| Xu[37] | 2.23 | 0.92 | 44.2 | 1.66 | 2.21 | 53.5 | 1.423 | 0.34 | 0.171 | 1.42 | 0.66 | 0.202 |
| Linnick[38] | 2.16 | 0.93 | 43.9 | 1.54 | 2.28 | 53.6 | 1.38 | 0.337 | 0.169 | 1.37 | 0.7 | 0.199 |
| Present | 2.13 | 0.95 | 44.1 | 1.572 | 2.30 | 53.2 | 1.386 | 0.349 | 0.166 | 1.383 | 0.70 | 0.195 |

The streamlines near the circular cylinder are depicted in Fig. 6 for Re = 20, even though the inner fluid is fictitious in physics. It is clearly seen no unphysical streamline-penetration appears, and the inner fluid cannot escape from the cylinder. Those results indicate that the no-slip boundary condition is accurately enforced in the present method. The recirculation length $L_w$, separation angle $\theta$, drag and lift coefficients ($C_d$ and $C_l$), and Strouhal number $St$ are presented in Table 1, including the data available in the literature [36-38] for comparison. Good agreement can be found between the present and the reference results.

## 3.2. Oscillating circular cylinder in a quiescent fluid

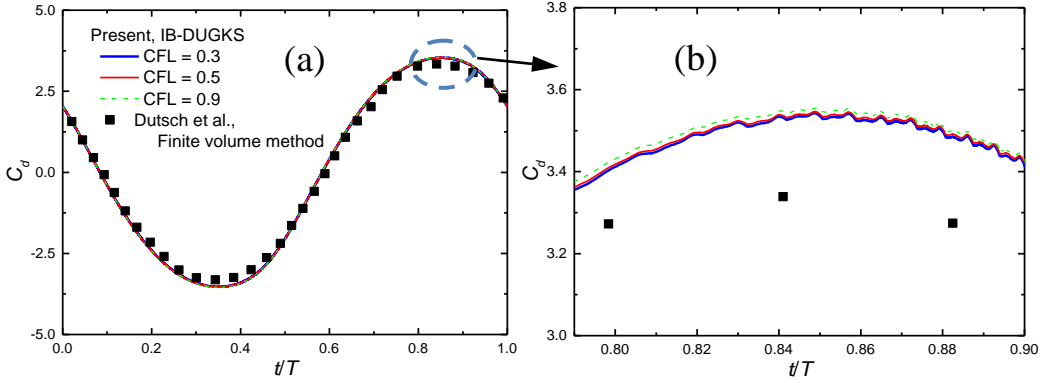

**Fig. 7.** Drag coefficient of a cylinder oscillating in a fluid during a cycle with different CFL numbers (a). (b) is the zoom-in view of (a).

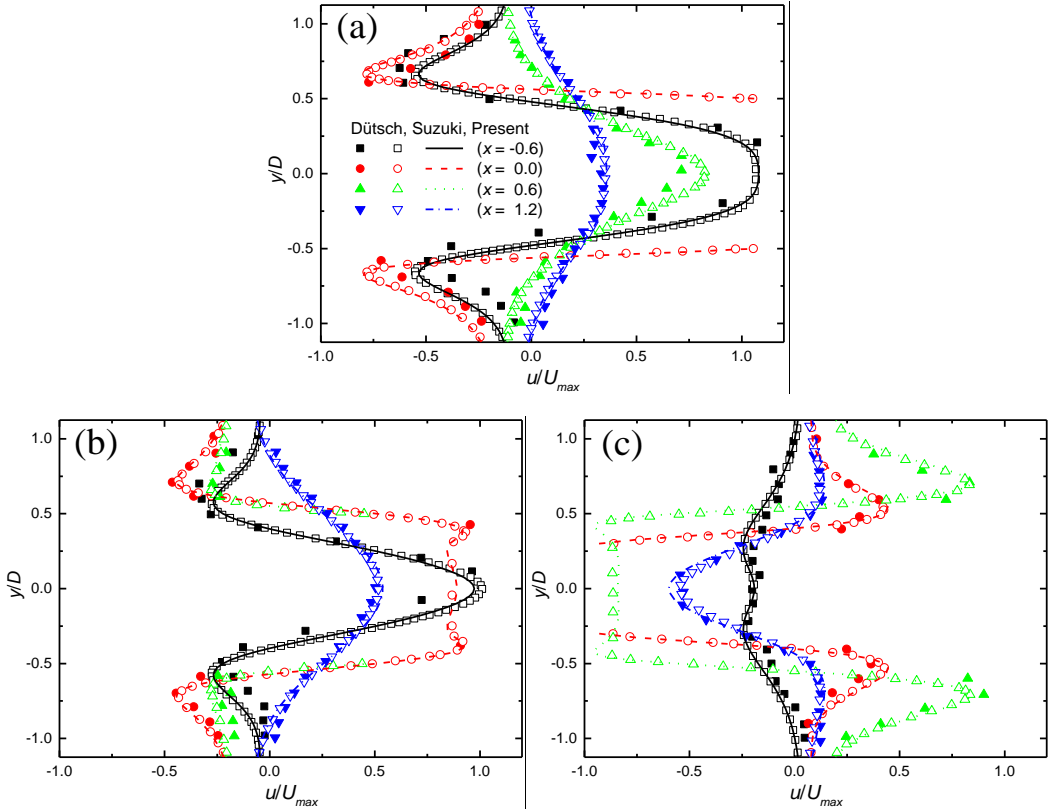

**Fig. 8.** Velocity profiles of the flow around an oscillating cylinder at four locations $x = -0.6 D$, 0, 0.6 $D$ and 1.2 $D$ at three phase angles $\phi = 180°$((a) $u/U_{max}$, $210°$((b) $u/U_{max}$) and $330°$((c)).

To explore the capability of the present IB-DUGKS in handling moving bodies, the flow induced by a circular cylinder oscillating in the stationary fluid is investigated. This case has been studied experimentally and numerically by Dütsch et al. [40] and Suzuki et al. [35]. The computational size in our simulations is $40D \times 30D$, with $D$ being the diameter of the cylinder locating at the center of the domain initially. The cylinder moves periodically in the $x$-direction with the motion equation of its center as $x(t) = -A\sin(2\pi f_c t)$, where $A$ and $f_c$ are the amplitude and frequency, respectively. The maximum velocity of the cylinder is $U_{max} = 2\pi f_c A$. The flow is mainly controlled by the Reynolds number, $Re = \rho_f U_{max} D/\mu$ and the Keulegan-Carpenter number, $KC = U_{max}/f_c D$. In the simulations, we set that $Re = 100$ and $KC = 5$. The Neumann boundary condition of $\boldsymbol{n} \cdot \nabla f = 0$ ($\boldsymbol{n}$ the unit vector normal to the boundary) is applied to the outer boundaries of the domain. In the simulations, the particle is resolved by 40 grids. An uniform mesh system ($1600 \times 1200$) for the computational domain is used.

We first examine the influence of the CFL number on the simulation results. Figure 7 shows the drag coefficient $C_d$ of the oscillating cylinder in a cycle $T = 1.0/f_c$ with three CFL numbers, i.e., 0.3, 0.5 and 0.9. It is observed that all the numerical results agree well with the data from Dütsch et al. [40]. This indicates the CFL number has little influence on the present results. The velocity profiles in the vertical cross-section at four locations $x = -0.6D$, $0$, $0.6D$ and $1.2D$, and for three phase angles $\phi = 180°$, $210°$ and $330°$ are presented in Fig. 8, where the coordinate is relative to the equilibrium position of the cylinder ($20D$, $15D$), and $\phi = f_c t \times 360°$. It can be found that the present results have a good agreement with the experiment and numerical results [35,40].

### 3.3. A single particle settling in channel

In this section, the flows with a single freely moving particle are to be considered. The first case is the sedimentation of a two-dimensional circular particle under gravity in an open channel, which is well documented in numerical experiments. The width and height of the channel are $W$ and $H$, respectively. A particle with diameter $D$ is placed initially at the center of channel. It is known that the characteristic of circular particle sedimentation is determined by the Reynolds number $Re = \mu u_s D/\rho_f$ and the ratio of width $\overline{W} = W/D$, where $u_s$ is the terminal velocity of particle. $\mu$ and $\rho_f$ are the dynamic viscosity and density of fluid, respectively. For small $Re$ and large $\overline{W}$, the drag force on a circular particle at steady state can be approximated as

$$\boldsymbol{F}_d = \frac{1}{4}\pi D^2 \left(\rho_s - \rho_f\right)\boldsymbol{g} = 4\pi K \mu \boldsymbol{u}_s, \tag{50}$$

where $\rho_s$ is the particle density. The parameter $K$ is the correction factor representing the hindering effect of the channel walls, which is a function of $\overline{W}$,

$$K = \left(\ln\overline{W} - 0.9157 + 1.7244\overline{W}^{-2} - 1.7302\overline{W}^{-4} + 2.4056\overline{W}^{-6} - 4.5913\overline{W}^{-8}\right)^{-1}. \tag{51}$$

The terminal settling velocity of the particle can then be derived from Eq. (50) as

$$\boldsymbol{u}_s = \frac{D^2}{16K\nu}\left(1 - \frac{\rho_s}{\rho_f}\right)\boldsymbol{g}. \tag{52}$$

In the simulation, the channel width is $W = 5D$ with $D = 0.24$ cm. The particle is resolved by 24 grids per diameter. The kinematic viscosity is set to be $0.1$ cm$^2$/s. Three sets of particle-fluid density ratios are considered, $\rho_r = \rho_s/\rho_f = 1.01$, $1.02$ and $1.05$. Those computational parameters are taken the same as those in [41]. We set the channel height $H = 80D$, which is long enough such that the inlet and outlet boundaries have no-visible influence on the simulation results. The fluid velocity at the inlet of channel is assumed to be zero, and at the channel outlet free-stream boundary condition is applied. No-slip boundary conditions are applied to the channel walls.

The settling velocity of the particle with time is shown in Fig. 9, together with the numerical results by Nie *et al.* [41] and Wang et al. [42], and the analytical solutions given by the Eq. (52). It is clearly seen that the present results agree well with those in [41,42]. Table 2 presents the terminal Re numbers correspondingly to the three density ratios. Compared to those of Wang et al. [42], the present results fit more closely to the data of Nie et al. [41]. Note that the numerical results all deviate from the analytical solutions at $\rho_r = 1.05$, which was also found in [41,42]. This can be ascribed to the failure of Eq. (52) for high Re [41].

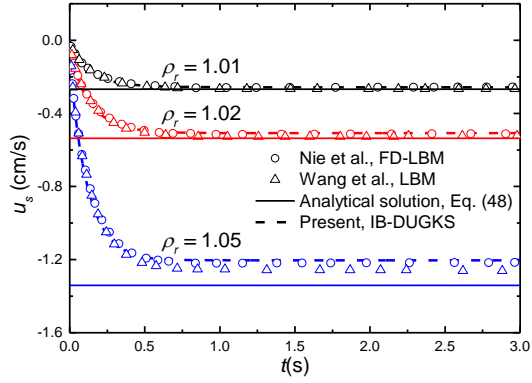

**Fig. 9.** Time history of the settling velocity of the particle with different density ratios.

TABLE 2. The terminal Reynolds number of the particle.

| $\rho_s/\rho_f$ | 1.01 | 1.02 | 1.05 |
|---|---|---|---|
| Nie et al. | 0.63 | 1.24 | 2.92 |
| Wang et al. | 0.64 | 1.27 | 3.02 |
| Present | 0.63 | 1.23 | 2.90 |

Secondly, to examine the reliability of the present IB-DUGKS for irregular particle suspension flows, the sedimentation of an elliptical particle is then simulated. Since an elliptical particle is anisotropic in shape, the particle-fluid interaction is expected to be more complex. For this problem, the numerical result from Xia et al. [43] using the finite element method is taken as the benchmark solution. The results obtained by the LBM with corrected moment exchange (CME) given in Chen et al. [44] and Caiazzo et al. [45] will also be used for comparison. The layout of this test case is depicted in Fig. 10. An ellipse with the major axis $a$ and minor axis $b$ is placed at the center of the channel, where the channel height $H$ and width $W$ are 12 and 0.4 cm, and $a = 0.05$ cm. The aspect ratio $a/b$ and the initial orientation angle $\theta$ of the particle are set to be 2 and $\pi/4$, respectively. The density of the particle is $\rho_s = 1.1$ g/cm$^3$ and the fluid viscosity is $\nu = 0.01$ cm$^2$/s. The computational sets mentioned above are the same as those in the previous studies [43,44]. The initial distance of the particle is 2.0 cm from the upper end of channel, and the particle is resolved by 40 grids along the major axis.

Figure 11 shows the trajectory and orientation of the particle during the settling process. It can be seen clearly that the present results are in good agreement with those obtained by other numerical methods [43,44]. Particularly, the ellipse has a trend of moving to the center of channel again, and the major axis is perpendicular to the centerline of channel at the steady state. This phenomenon is consistent with that reported in [43]. To examine the accuracy of the present IB-DUGKS in the force calculation, the hydrodynamic force exerted on the particle at a short period of time (forty temporal steps) is presented in Fig. 12, together with those obtained by the LBM with CME schemes [44,45]. It is observed that even though the average values of the three appears comparable, some fluctuations are found in the results by Chen et al. [44] and Caiazzo et al. [45], which can be attributed to unphysical oscillations in the pressure field inherent in LBM [34] due to some inconsistence at the point, moving out the solid body and vice versa. On the other hand, the force in the present IB-DUGKS is quite smooth.

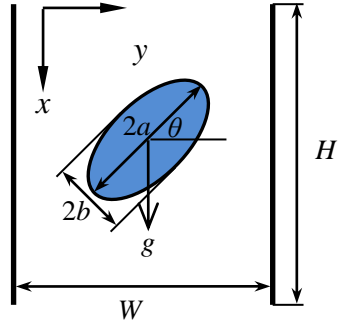

**Fig. 10.** The geometry of an elliptical particle settling in a channel.

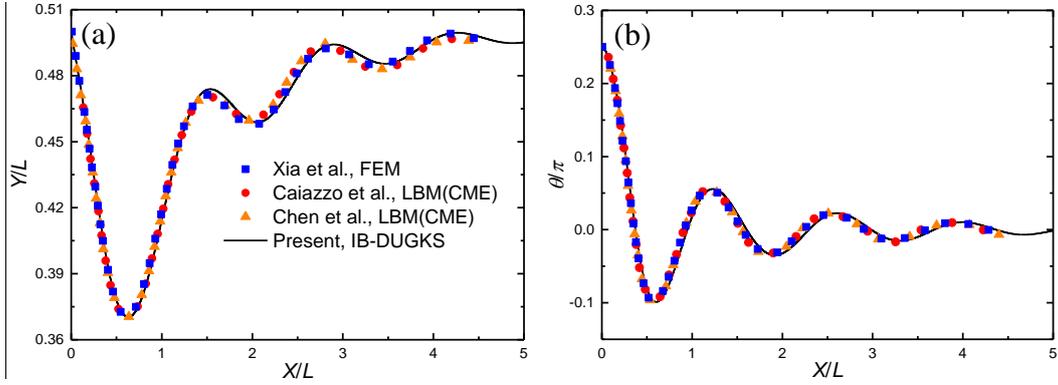

**Fig. 11.** The trajectory and orientation of the elliptical particle settling in a channel.

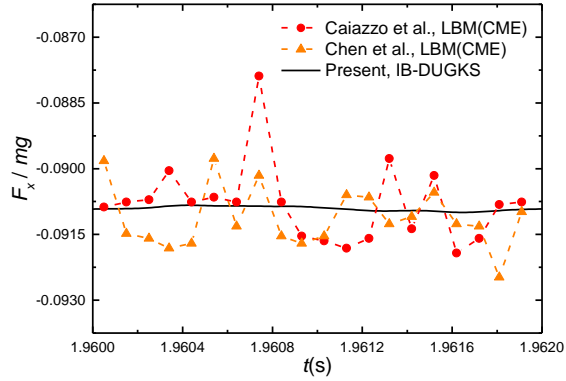

**Fig. 12.** Comparison of the fluid force in the *x*-direction when the particle declines steadily. *mg* is the gravity of the particle.

Thirdly, the simulation of a spherical particle settling under gravity in an enclosure is further conducted to test the proposed scheme for the three-dimensional particulate flows. This flow problem has been studied experimentally by Cate et al. [53] using the PIV technology and Feng et al. [18] through numerical method (IB-LBM). The size of the rectangular cavity is $10 \times 10 \times 16$ cm, which contains fluid with four sets of density and dynamic viscosity as $(\rho_f, \mu_f)$ = (0.97, 3.73), (0.965, 2.12), (0.962, 1.13) and (0.96, 0.58) (g/cm$^3$ g/(cm.s)), called case 1 to 4, respectively. According to the measurements of Cate et al. [53], the terminal particle Reynolds numbers for the four cases are 1.5, 4.1, 11.6 and 31.9. The diameter of the sphere is $D = 1.5$ cm, with a density fixed at $\rho_s = 1.12$ g/cm$^3$. The particle center is initially located at (5, 5, 12.75) cm. In the simulation, the grid system is $120 \times 120 \times 192$ for the computational domain.

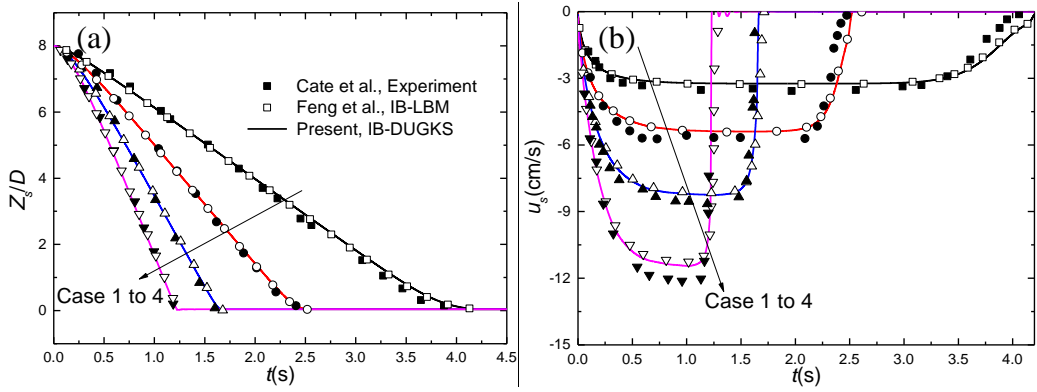

**Fig. 13.** The vertical position (a) and velocity (b) of a sphere settling under gravity.

Figure 13 presents the position (a) and velocity (b) of particle in the z-direction, including the data from Cate et al. [53] and Feng et al. [18] for comparison. It can be found that the present results agree reasonably well with those in the literature, especially for the lower particle Reynolds number cases. Also noticed is that a relative large velocity deviation emerges in case 4 with the highest particle Re. This phenomenon has been observed in the previous studies [18,47]. The reason may be attributed to the diffuse nature of the IB method.

### 3.4. DKT of two particles in channel

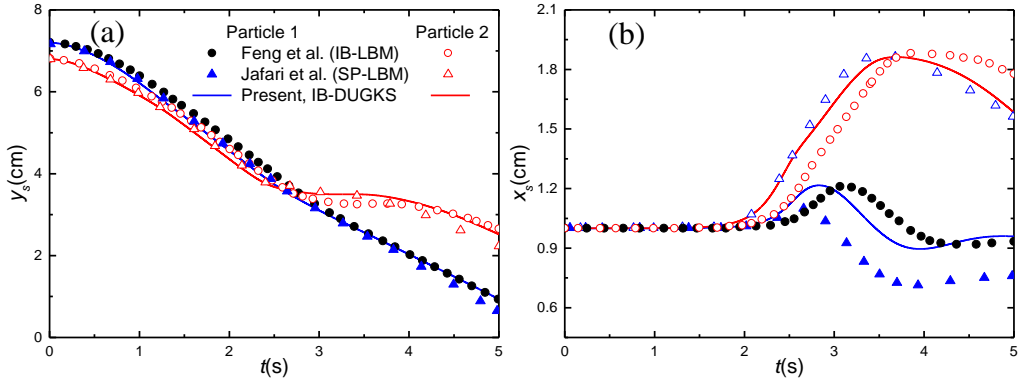

**Fig. 14.** The trajectory of the two particles in the $x$ (a) and $y$ (b) directions.

In actual particulate flows, a particle is usually not only interacting with the fluid, but also undergoes frequent inter-particle interactions. In this section, the simulation of a particle pair settling under gravity is performed to further evaluate the present IB-DUGKS in modeling multiple particle systems. This standard test case considered is two-dimensional, and has been used in many previous studies. In the present study, the parameters are chosen the same as those in [12,46,47]. The size of the channel is 2 cm × 8cm with $D = 0.2$ cm being the particle diameter, and the kinematic viscosity of fluid is $\nu = $ 0.01 g/(cm. s). The upper and lower particles are identical with the same density of $\rho_p = 1.01$ g/cm$^3$. The initial positions of the two particles are (0.999 cm, 7.2 cm) and (1.0 cm, 6.8 cm), respectively. In order to break the strong symmetry of the flow and to induce tumbling motion later on, a slight deviation in the $x$-direction is set intentionally for the first particle. In the simulations, the particle is resolved by 20 grids corresponding to a uniform mesh with size of 200 × 800 for the computational domain, and the CFL number is set to be 0.5.

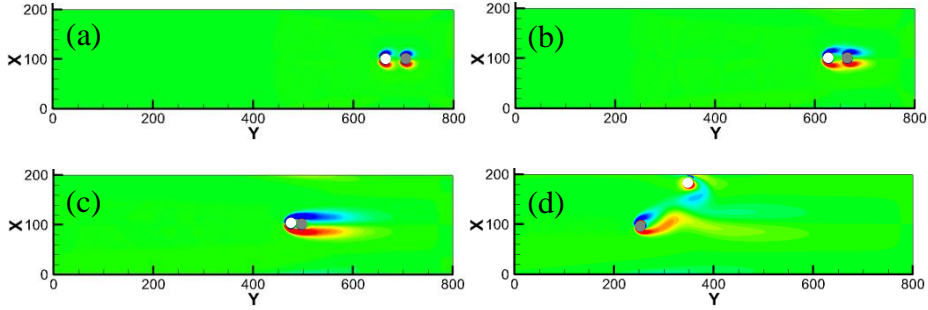

**Fig. 15.** The instantaneous vorticity at $t = 0.354$ (a), 0.708 (b), 1.77 (c), 3.54 s (d) during the settling process of two particles in a channel.

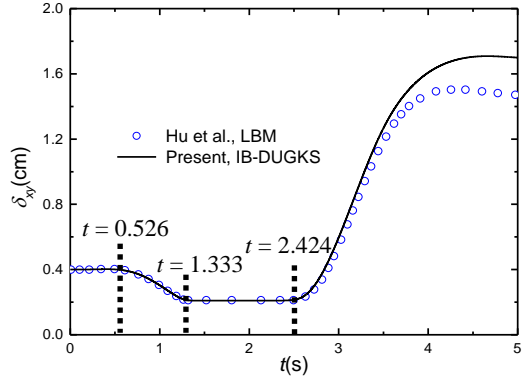

**Fig. 16.** Time history of the distance between the two particles.

Figure 14 presents the instantaneous positions of the two particles during the settling process. The DKT phenomenon is clearly reproduced. Initially, the two particles are located along the centerline of channel with a relatively small gap. After being released from rest in the still fluid, both particles begin to descend under gravity, as shown in Fig. 15(a). While the leading particle is falling down, it creates a wake with lower pressure. As the trailing particle comes close to the leading one, it is drafted into the wake and experiences a much smaller drag (Fig. 15(b)). Hence, the trailing particle moves faster than the leading one, and eventually catches up, and then kisses and impels the latter. This stage persists for some time, during which the particles form a doublet and fall downwards together (Fig. 15(c)). However, that state is unstable as indicated in [12,46], because of some symmetry breakings such as the fluctuating wake. As a result, the sedimentation process turns into the tumbling stage, where the particles start to separate from each other (Fig. 15(d)). The time history of the center to center distance of the particles is given in Fig. 16. It can be seen that after about 0.526 seconds, the gap decreases gradually. At about $t = 1.333$ s, the distance approaches to a local minimum value, indicating a contact with each other, and this kissing stage lasts about 1.091 seconds. Finally, at about $t = 2.424$s, the distance increases and the particles start to separate from each other. As shown in Fig. 16, the DKT processes predicted by the IB-DUGKS agree well with those reported by Hu et al. [47] using the LBM. However, noticeable differences in the tumbling stages are seen for the two results. As indicated in [46,47], the difference can be expected in that the dynamics in the tumbling phase relies heavily on the growth rate of the numerical uncertainties and the boundary treatments as well as the collision models. Hence, the present IB-DUGKS can be generally considered to be able to produce reasonable results for the DKT dynamics of two particles.

### 3.5. A group of particles settling in an enclosure

In order to further explore the capability of the present IB-DUGKS for more complex particle suspension flows, simulations of a group of particles freely moving in an enclosure are conducted. Note that the particles are equal in size. The first problem considered for confirming the stability of the present scheme is the flotation of 5 three-dimensional spheres in an enclosure. The cavity has a size of $10 \times 10 \times 25$ cm. In the simulations, the fluid viscosity is set to be 2.0 cm$^2$/s, and the sphere diameter is 1.5 cm. The density ratio of particle to fluid is fixed at 0.8, indicating lightweight particles in fluid. The initial positions of the centers of first sphere is (5, 5, 1.25), surrounded by other four spheres (2 to 5) located at (4, 4, 3.75), (6, 4, 3.75), (4, 6, 3.75) and (6, 6, 3.75) cm, respectively. The sphere diameter is resolved by 12 grids.

Figure 17 presents the snapshots of the five spheres at times $t = 0$, 0.5, 1.0, 2.0 and 3.6 s. The spheres are floating upwards, since its density is lower than that of the ambient fluid. What's more, it is found that sphere 1 moves faster than the others as the distance between them decreases at $t = 0.5$ s. The reason is that the trailing particle (sphere 1) settles in the low-pressure wake of the leading spheres, and then its drag coming from the fluid reduces. This is similar to the case of two circular particles settling under gravity, shown in Fig. 15. With the time increasing, sphere 1 catches up with the others ($t = 1.0$ s), and then crosses through the four spheres ($t = 2.0$ s), pushing the other spheres aside simultaneously. Eventually, sphere 1 first reaches the upper wall ($t = 3.6$ s). The vertical velocity and position of spheres 1 and 2, and the trajectory of spheres 1 to 3 are given in Fig. 18. The results from Yasushi et al. [51] using the iSP-LBM are also included for comparison. The results are consistent with those shown in Fig. 17, and good agreement can be found by the simulations of the present IB-DUGKS and iSP-LBM from Yasushi et al. [51].

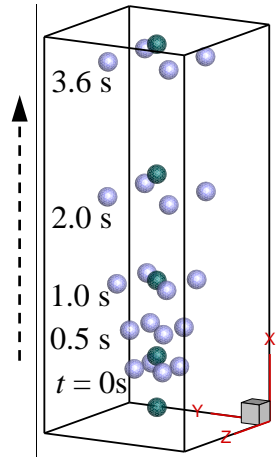

**Fig. 17.** The flotation process of five spheres in a closed cavity.

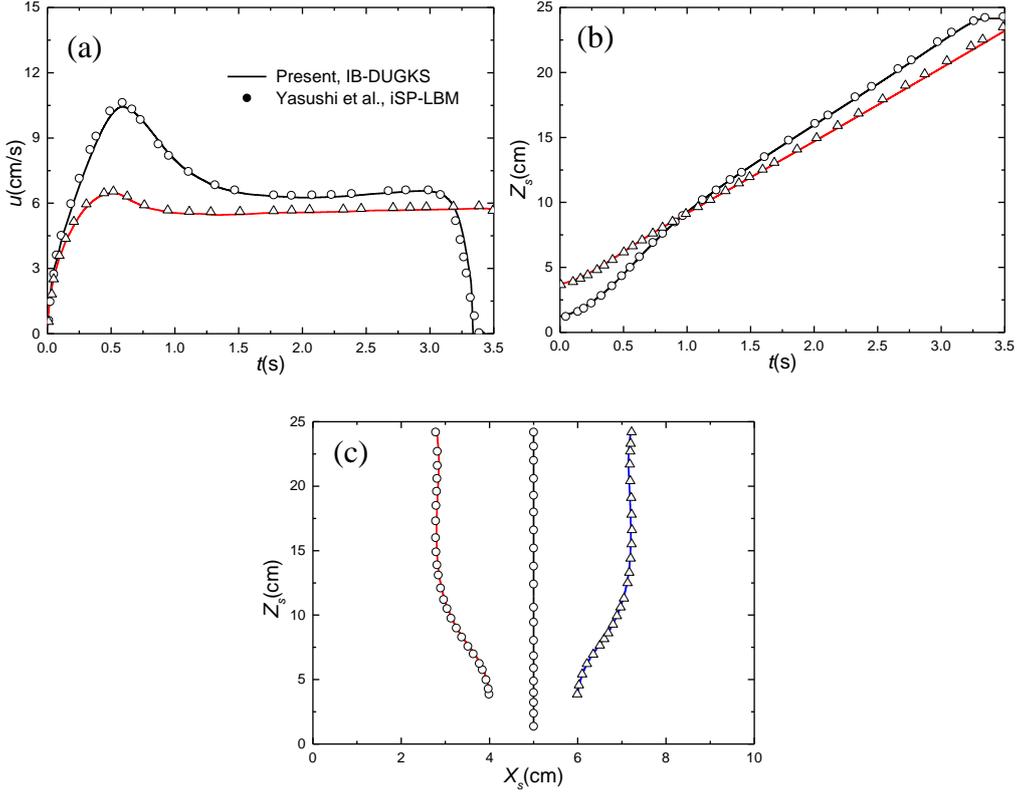

**Fig. 18.** Time histories of the vertical velocity (a) and position (b) of spheres 1 and 2, and the X-Z trajectory of spheres 1 to 3 (c).

The simulation of a group of circular particles settling in an enclosure is further conducted. This problem has been extensively studied using the finite element method [48], lattice Boltzmann method [12] and fictitious domain method [41]. In this case, the system dynamics is governed by both particle-particle and particle-vortex interactions, resulting in the emergence of instability of particle cluster during the sedimentation process.

In the simulation, 128 particles with diameter $D = 0.1$ cm and density $\rho_s = 1.01 \text{g/cm}^3$ are initially aligned uniformly at the top of a 2 cm $\times$ 2 cm square cavity, which is filled with a fluid with kinematic viscosity $v = 0.01 \text{ cm}^2/\text{s}$. The particles resolved by 20 grids are then released with zero velocity and fall downwards under the gravity. When particle-particle and/or particle-wall contacts occur, the collision model described in section 3.2 will be imposed. The boundaries of the cavity in both directions are static walls, which are realized by the simple bounce back rule. The CFL number is set to be 0.5 in the simulations.

The snapshots of the evolution process and the vorticity field are shown in Fig. 19. The typical feature of this problem, i.e., the development of the Rayleigh-Taylor instability can be clearly observed in Figs. 19(b) to 19(d). At the beginning of sedimentation, the particles fall almost uniformly (Fig. 19(b)), except for the two lateral columns of particles closest to the cavity walls. This can be attributed to the hindering effect of wall that produces the vortices at two sides of the advancing front (Fig. 19(c)). Meanwhile, instability generates in the squeezed out particles, as the particles near the sidewalls drop quickly while those in the middle are blocked by the fluid. Then, an umbrella-shaped structure emerges, as shown in Fig. 19(d). The findings are similar to those reported in [12,48,41]. Eventually, the particles start to reach the bottom of the cavity, and interestingly a bubble is formed in the center of the lower half of the cavity (Fig. 19(e)). However, the bubble structure collapses in that some stronger eddies push the particles up to the fluid again (Fig. 19(f)). The particles are well blended in this stage. After that, the packing process is initiated (Fig. 19(g)), and finally all particles settle and are packed on the bottom wall, and the fluid gradually returns to rest (Fig. 19(h,i)). We can observe that the complex dynamics during the sedimentation process can be successfully predicted by the present IB-DUGKS.

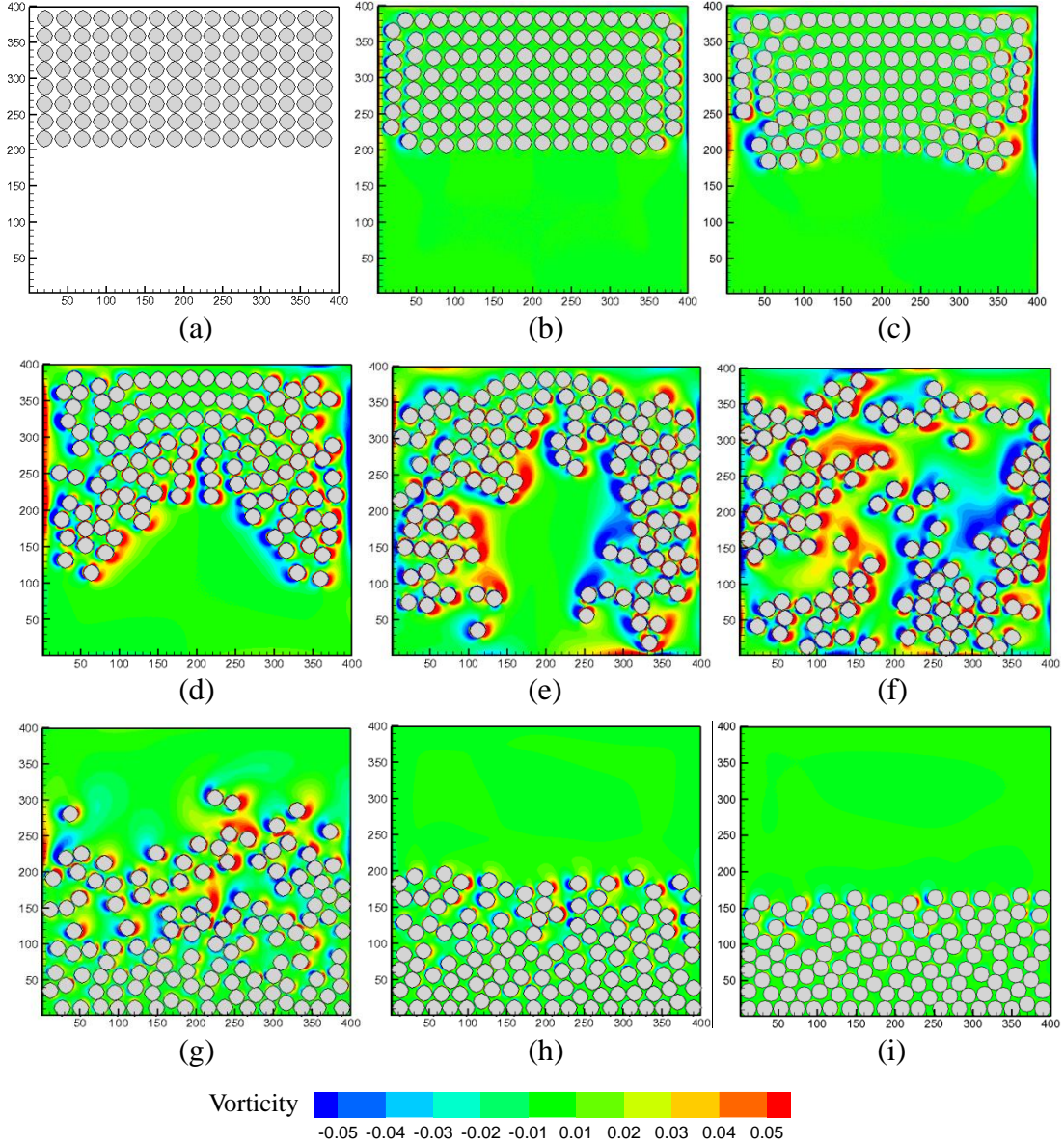

**Fig. 19.** The sedimentation process of 128 circular particles in an enclosure. (a) to (i): $t = 0, 3, 5, 8, 10, 12, 20, 30$ and $36$ s, respectively.

## 4. Conclusions

In this work, an immersed boundary-discrete unified gas kinetic scheme (IB-DUGKS) is developed to allow interface-resolved simulations of the two- and three-dimensional particulate flows. The present method solves the flow field using the DUGKS, which is a finite volume scheme based on the Boltzmann-BGK equation. The fluid-solid interface is transformed to a virtual force field using the immersed boundary method. This non-uniform force term is conveniently incorporated into the DUGKS by the Strang-Splitting algorithm, in which the IB force only influences the distribution functions at the cell center. The no-slip boundary condition on the particle surface is accurately enforced by an iterative algorithm, so that the drawback of the unphysical penetration of streamlines in the conventional IBM can be fully removed.

Simulations of several flows with stationary and moving cylinders are carried out to test the accuracy of the present IB-DUGKS, and three particle-laden flows are implemented to further validate the feasibility of the IB-DUGKS, including the sedimentations of a circular and an elliptical particle in a channel, the drafting- kissing-tumbling dynamics of two settling particles, and the sedimentation of a cluster of particles in an enclosure. The results are found to be in good agreement with the analytical, experimental and numerical data available in the literature. These tests together illustrate the reliability and flexibility of the present IB-DUGKS for simulating the particulate flows.

## Acknowledgments

The authors acknowledge the support by the National Natural Science Foundation of China (51390494) and the U.S. National Science Foundation under grants CNS-1513031, CBET-1235974, and AGS-1139743.

## Appendix: the update of $\widetilde{f}$ in the Strang-Splitting scheme

Taking the pre-forcing step for example, the original distribution function and macroscopic quantities are calculated as

$$f^* = f^n + 0.5\Delta t \boldsymbol{F}(\rho^n, \boldsymbol{u}^n), \tag{1}$$

$$\rho^* = \rho^n, \quad \boldsymbol{u}^* = \boldsymbol{u}^n + 0.5\boldsymbol{a}\Delta t \tag{2}$$

In terms of $\tilde{f}$ actually tracked in the DUGKS, we have

$$\tilde{f} = \frac{2\tau + \Delta t}{2\tau} f - \frac{\Delta t}{2\tau} f^{eq}(\rho, \boldsymbol{u}), \tag{3a}$$

$$\tilde{f}^* = \frac{2\tau + \Delta t}{2\tau} f^* - \frac{\Delta t}{2\tau} f^{eq}(\rho^*, \boldsymbol{u}^*). \tag{3b}$$

From Eqs. (3a), (3b) and (1), we can obtain

$$\begin{aligned}\tilde{f}^* &= \tilde{f} + \frac{2\tau + \Delta t}{2\tau}\left(f^* - f\right) - \frac{\Delta t}{2\tau}\left(f^{eq}(\rho^*, \boldsymbol{u}^*) - f^{eq}(\rho, \boldsymbol{u})\right). \\ &= \tilde{f} + \frac{(2\tau + \Delta t)\Delta t}{4\tau}\boldsymbol{F}(\rho^n, \boldsymbol{u}^n) - \frac{\Delta t}{2\tau}\left(f^{eq}(\rho^*, \boldsymbol{u}^*) - f^{eq}(\rho, \boldsymbol{u})\right)\end{aligned} \tag{4}$$

Note that

$$\nabla_u f^{eq} = \frac{(\boldsymbol{\xi} - \boldsymbol{u})}{RT} f^{eq} = -\nabla_{\boldsymbol{\xi}} f^{eq}. \tag{5}$$

Then using Taylor expansion and the results of Eq. (31) in the main text, we can approximate $f^{eq}(\rho^*, \boldsymbol{u}^*)$ as

$$\begin{aligned}f^{eq}(\rho^*,\boldsymbol{u}^*) &= f^{eq}(\rho,\boldsymbol{u}^*) = f^{eq}(\rho,\boldsymbol{u}+0.5\boldsymbol{a}\Delta t) \approx f^{eq}(\rho,\boldsymbol{u})+0.5\Delta t\boldsymbol{a}.\nabla_u f^{eq}(\rho,\boldsymbol{u})\\ &= f^{eq}(\rho,\boldsymbol{u})-0.5\Delta t\boldsymbol{a}.\nabla_\xi f^{eq}(\rho,\boldsymbol{u}) = f^{eq}(\rho,\boldsymbol{u})+0.5\Delta t\boldsymbol{F}(\rho^n,\boldsymbol{u}^n).\end{aligned} \qquad (6)$$

Substituting Eq. (6) into Eq. (4), we have

$$f^* = f + \frac{(2\tau+\Delta t)\Delta t}{4\tau}\boldsymbol{F}(\rho^n,\boldsymbol{u}^n) - \frac{\Delta t}{2\tau}\big(0.5\Delta t\boldsymbol{F}(\rho^n,\boldsymbol{u}^n)\big) = f + 0.5\Delta t\boldsymbol{F}(\rho^n,\boldsymbol{u}^n) \qquad (7)$$